\documentclass[prb,aps,twocolumn,groupedaddress,floatfix]{revtex4-1} 

\usepackage{amsfonts}\usepackage{amssymb}
\usepackage{amsmath}\usepackage{graphicx}
\usepackage{bbold}\usepackage{textcomp}
\usepackage{color,soul}
\usepackage[normalem]{ulem}\usepackage{MnSymbol}
\usepackage{epsfig}\usepackage{bm}
 \usepackage{times}
  \usepackage{lineno}
\newcommand{\upmu}{\mu}

\begin{document}
\title{Backaction in metasurface etalons} 

\author{Andrej Kwadrin}
\affiliation{Center for Nanophotonics, FOM Institute AMOLF, Science Park 102, 1098XG Amsterdam, The Netherlands}
\author{Clara I. Osorio}
\affiliation{Center for Nanophotonics, FOM Institute AMOLF, Science Park 102, 1098XG Amsterdam, The Netherlands}
\author{A. Femius Koenderink}
\email{Corresponding author: f.koenderink@amolf.nl}
\affiliation{Center for Nanophotonics, FOM Institute AMOLF, Science Park 102, 1098XG Amsterdam, The Netherlands}


\begin{abstract}
We consider the response of etalons created by a combination of a conventional mirror and a  metasurface composed of a periodic lattice of metal scatterers with a resonant response. This geometry has been used previously for hybridization of localized plasmons with Fabry-Perot resonances and for perfect absorption, in so-called Salisbury screens. The particular aspect we address is if one can assume an environment-independent reflectivity for the metasurface when calculating the  reflectivity of the composite system, as in a standard Fabry-Perot analysis, or whether the fact that the metasurface interacts with its own mirror image renormalizes its response. Using lattice sum theory, we take into account all possible retarded dipole-dipole interactions of scatterers in the metasurface amongst each other, and through the mirror. We show that while a layer-by-layer Fabry-Perot formalism captures the main qualitative features of metasurface etalons, in fact  the mirror modifies both the polarizability and reflectivity of the metasurface in a fashion that is akin to Drexhage's modification of the radiative properties of a single dipole.
\end{abstract}


\maketitle
\section{Introduction}

Local enhancement of electromagnetic fields by periodic lattices of metal scatterers has many applications, which include improving refractive index sensors~\cite{Altug2010,zijlstra2012}, brightening solid-state light emitters~\cite{zayatsled,lozano2013}, and providing substrates for linear and nonlinear optical spectroscopies~\cite{giannini2011}. These applications on one hand employ the fact that  single metal objects can provide strong scattering resonances on basis of plasmonic (i.e., free electron) oscillations. In addition,  these objects can be arrayed either in a diffractive ``plasmonic crystal''\cite{Zou2004,lozano2013} or can be packed more densely in a ``metasurface'' or ``frequency-selective'' surface~\cite{FSSreview}. Several works have investigated how the response of such periodic plasmonic antenna arrays can be further modified by placing them close to a ``ground plane'', i.e., a reflective interface.  In fact, this idea dates to the 1950's when the ``Salisbury screen'' was proposed as a highly efficient microwave absorber in radar applications~\cite{salisbury52}. This absorber is created by placing a weakly reflective and slightly absorptive layer at a quarter wavelength from a mirror. The concept has recently been extended, for instance, to mid-IR absorption in graphene~\cite{AtwaterSalisbury}, and, in an independent approach, to metamaterial arrays at $<\lambda/10$ distances from ground planes  to reach above 97\% absorption in the mid-IR and THz range~\cite{padilla08,padilla08b,padilla09,padilla10}.   Ameling and Giessen have also pursued plasmonic arrays near reflective interfaces  in an effort to enhance plasmonic sensors by strongly coupling localized plasmons with Fabry-Perot resonances~\cite{Ameling10,Ameling10b,Ameling11,AmelingLPRev2013}. These approaches combine two essential ingredients:  (1)  a resonant scattering layer that provides a reflection constant going through a 180$^\circ$ phase shift and an amplitude maximum, and (2) tuning of a  geometrical resonance associated with the distance to one, or two, reflectors.

While the structures outlined above have been the subject of ample numerical simulations, we have found it instructive to re-analyze their physics  from two different perspectives.  First, we note that the geometry of an etalon with a weakly reflective interface in front of a unit reflectivity interface is of interest not just as an absorber, but also for its phase properties.  In particular, this geometry, known as a Gires-Tournois etalon, provides large chromatic dispersion even with a non-resonant front interface~\cite{girestournois}. Here we examine the phase response taking a resonant  metasurface as front interface. Second, in addition to reporting on the phase behavior of metasurface etalons, we are interested in determining how strongly the metasurface scatterers become polarized, and what the role of backaction is. It has been previously assumed that it is possible to  assign a dispersive reflection constant to the bare metasurface, to then apply a Fabry-Perot multilayer reflectivity formalism to predict the reflectivity of a composite system. This assumption is difficult to reconcile with the established notion of backaction, whereby the scattering strength and linewidth of plasmonic scatterers can be  modified by reflective interfaces due to interference of an induced dipole with its own mirror image~\cite{Buchler2005,Kwadrin2013PRB,Drexhage1968,Drexhage1970}.  

We report on a self consistent lattice sum theory that includes backaction, as well as all dipole-dipole interactions of scatterers in the metasurface, both directly~\cite{Lunnemann2013,GarciadeAbajo2007} and through multiple scattering off the nearby reflective interface.  This serves as benchmark for the Fabry-Perot multilayer reflectivity model postulated by Ameling et al.~\cite{AmelingLPRev2013}.  We find that the Fabry-Perot model indeed gives an excellent estimate of the composite etalon performance. In particular,  for dense lattices (pitch $\lesssim\lambda/5$) and large distances from lattice to interface ($\gtrsim\lambda/5$), the model gives quantitatively accurate results.  For more dilute lattices,  the Fabry-Perot model  reproduces all qualitative features, but reflection phase and amplitude can be off by as much as $\sim 0.5$ rad and $0.1$ respectively. These deviations show the effect of backaction, or equivalently, of multiple scattering interactions with the mirror via evanescent grating orders. In Fabry-Perot terms,  the equivalent metasurface reflection constant is \emph{not} independent of its environment. This result indicates that the response of stratified metasurface systems cannot be rigorously captured in an effective multilayer formalism, even if in particular practical cases the multilayer formalism is appropriate.

\section{Fabry-Perot analysis  of metasurface etalons}
To assess the general performance of etalons with a  mirror and a metasurface interface, we first set up a simple two-interface model, similar to the efforts by Ameling et al.~\cite{Ameling10,AmelingLPRev2013}. We consider incidence from medium 1, onto a metasurface that is separated by a distance $d$ from a highly reflective interface (medium between the two reflective planes labeled 2, beyond that, medium 3, refractive indices denoted as $n_i$.). We denote with $k=\omega/c$ the free space wavenumber, and indicate with $r_{nm},t_{nm}$ the reflection and transmission amplitude coefficients when going from medium $n$ to $m$. The reflectivity of the stack, under the assumption that a  simple Fabry-Perot model is applicable, $r^{\mathrm{FP}}_{\mathrm{stack}}$, is given by  
\begin{equation}
r^{\mathrm{FP}}_{\mathrm{stack}}=r_{\mathrm{12}}+\frac{t_{\mathrm{21}}r_{\mathrm{23}}t_{\mathrm{12}}\left(-e^{2in_2kd}\right)}{1-r_{\mathrm{21}}r_{\mathrm{23}}\left(-e^{2in_2kd}\right)}
\label{eq:FPmodel}
\end{equation}
where $r_{23}$ is the reflectivity of the back reflector, taken from Fresnel equations for the reflectivity of a real metal mirror, while $r_\mathrm{12}$,  $r_\mathrm{21}$ and  $t_\mathrm{12}$, $t_\mathrm{21}$ refer to the metasurface.   Garc{\'i}a de Abajo \cite{GarciadeAbajo2005,GarciadeAbajo2007} has worked out the reflectivity of a periodic surface of scatterers in free space, under normal incidence as $r=r_\mathrm{12}=r_\mathrm{21}$, $t=t_\mathrm{12}=t_\mathrm{21}$, 
\begin{equation}
r=-\frac{2\pi i k}{a^2}\frac{1}{1/\alpha_\mathrm{free}-{\cal G}} \qquad \mbox{with} \qquad t=1+r
\end{equation}
with $\alpha_\mathrm{free}$ the particle polarizability (for a single scatterer, in absence of the interface), $a^2$ the area of the unit cell, and ${\cal G}$ an Ewald lattice sum that takes into account all dipole interactions. We refer to Ref.~\onlinecite{GarciadeAbajo2007} and the Appendix \ref{sec:green_app} for the definition and evaluation of free space lattice sums. Let us for simplicity assume plasmon nanobars, that is, particles  that are only polarizable along  one of their axis ($x$ as the incident polarization) and along one axis of the lattice, which applies for both  the experiments of Ameling et al.~\cite{Ameling11}, and the experiment described in Appendix \ref{Experiment}  (performed with split rings with negligible magnetic moment and for which, at normal incidence, the magnetic dipole does not contribute to reflection and scattering because is out of plane).  According to Eqs. (10) and (13) of Ref.  ~\onlinecite{GarciadeAbajo2007}, for cubic, nondiffractive lattices with $\lambda \geq a$ the sum is well approximated by 
\begin{equation}
{\cal G}=\left(\frac{4\pi^2\sqrt{2}}{a^3}\frac{1}{\sqrt{2\pi/(ka)-1}}-118\right) - i\left[\frac{2\pi k}{a^2} - \frac{2}{3}k^3\right].
\end{equation}
If the electrostatic polarizability $\alpha_0$ is chosen Lorentzian~\cite{Sersic2011PRB,deVries}, the full particle polarizability including radiation damping (also known as `dynamic polarizability') reads
\begin{equation}
\alpha_0=\frac{V\omega_0^2}{\omega_0^2 - \omega^2 - i\gamma\omega - i \frac{2}{3}k^3}
\end{equation}
with the term containing $k=n\omega/c$ representing radiation loss. Thus an explicit expression results for a resonant reflectivity around $\omega_0$ with Ohmic damping rate $\gamma$, and with a strength tuned through scatterer volume $V$. We discuss results for a typical scenario with scatterers resonant at telecom wavelengths ($1.25\times 10^{15}$~s$^{-1}$), taking $\gamma=0.1\omega_0$ typical of the Ohmic damping in gold. We choose a  volume $V=0.0020~\mu$m$^3$ to obtain an extinction cross section on resonance ($4\pi k \mathrm{Im}\alpha$)   halfway the maximum cross section for a strong dipole scatterer~\cite{GarciadeAbajo2007} ($3/2\pi \lambda^2$), as is typical for  plasmonic scatterer according to measurements in Ref.~\onlinecite{Husnik}.  To quantify $r_{23}$ we take the Fresnel coefficients for an air-silver interface, using tabulated optical constants for silver~\cite{Palik}.   We note that this Fabry-Perot model only uses normal-incidence reflectivities of the interfaces,  and neglects possible grating diffraction effects.  While in this paper we consider structures that have no far-field diffraction orders in the frequency range of interest,  the full model discussed in section~\ref{sec:full}  includes all diffraction orders.

\begin{figure}
\centering
\includegraphics[width=\columnwidth]{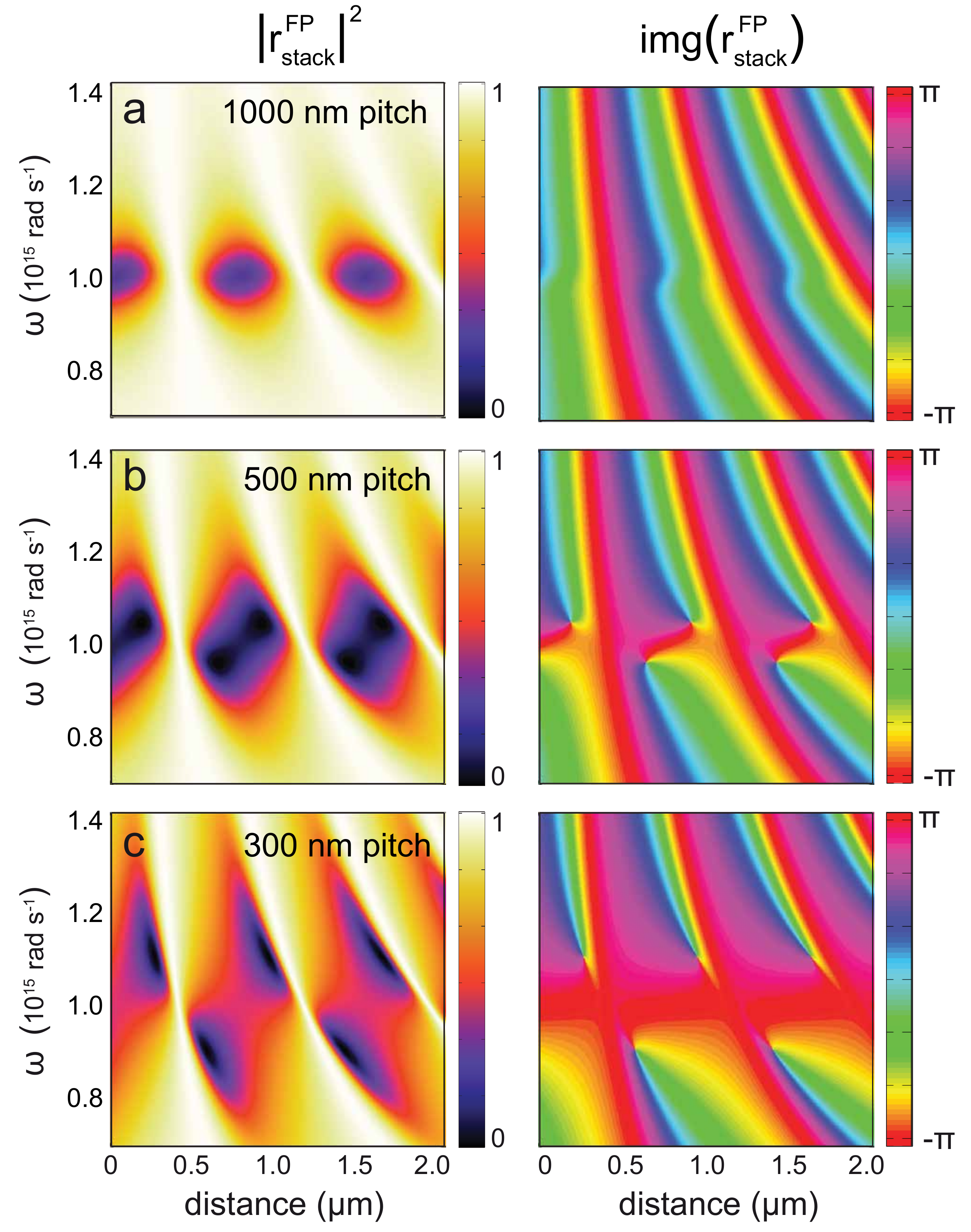}
\caption{Amplitude and phase of the stack reflectivity calculated using a simple Fabry-Perot model for dilute to dense systems. The pitches are indicated in the figure and the respective unit cell areas are   (a) $a^2=1$ $\mu m^2$, (b) $a^2=0.25$ $\mu m^2$  and (c) $a^2=0.09$ $\mu m^2$.}
\label{fig2:FPIprediction}
\end{figure}

Figure~\ref{fig2:FPIprediction} shows the expected amplitude and phase response for three cases ranging from a dense to a quite dilute lattice, where the bare lattice reflectivity goes from being much broader than $\gamma$ and strong   ($|r|^2>80\%$, with $<1\%$ transmission and 20\% absorption), to being limited in width to $\gamma$ and  weak ($|r|^2\sim 3\%$ at $|t|^2\sim 70\%$).  The etalon behavior is very different for these three cases. In the very dense case, the reflectivity shows very sharp bands of absorption around the integer roundtrip condition, with a lineshape that is asymmetric, and reverses asymmetry when $\omega$ sweeps through $\omega_0$. The very narrow slivers  of near-perfect absorption in $\omega-d$ space are associated with very sudden changes in phase of the reflectivity. As the lattice is diluted, the narrow stripes of absorption broaden in $d$-range and narrow in $\omega$-range.  Points of perfect absorption occur on either side of $\omega_0$, at opposite sides of the geometrical resonance condition, and are associated with phase singularities in the reflectivity. Finally, for the very dilute case, as function of spacing to the mirror pockets of near-perfect reflectivity and strong absorption alternate, while no phase singularities occur.  The following remarkable features stand out. First, perfect reflectivity can be obtained even though the single pass absorption by the metasurface is between 20\% and 50\% for the dense to dilute lattices.  Second, for the dense lattice the very broad and quite absorptive bare response for the lattice results in very sharp asymmetric response lines~\cite{AmelingLPRev2013}. Third, for the dilute lattices very strong reflection contrast is possible, even though the metasurface itself only has a few percent reflection.  In part this is understandable from the fact that the single pass absorption is quite high (30\% for the most dilute lattice).  This $k$ survey shows that a Fabry-Perot analysis of metasurface etalon behavior contains the physics of the Salisbury screen (highly reduced reflection and increased absorption), the Gires-Tournois etalon (strong chromatic dispersion even in wavelength ranges where $r_{12}$ does not change phase), and the apparent transition from crossing to anticrossing behavior of the Fabry-Perot dispersive resonance and the LSPR reported by Ameling et al.~\cite{salisbury52,girestournois,AmelingLPRev2013}

\section{Full theory\label{sec:full}}
Having introduced the physics of metasurface etalons in a Fabry-Perot model, we now turn to a full point-dipole model to confirm when the accepted Fabry-Perot intuition is correct, and to identify the origin of any deviations from it. Our goal is to find the full \emph{lattice-summed} Green function $\mathcal{G}$ for a system composed of a  infinite two-dimensional lattice of point scatterers at a distance $d$ from a parallel interface, taking into account all the multiple scattering interactions among particles, including those via the mirror. While we  only write down the theory for electric dipole moments $\mathbf{p}$,  the model directly extends to magnetic and magnetoelectric dipoles (read ($\mathbf{p},\mathbf{m}$) for $\mathbf{p}$ in all equations,  driven by $(\mathbf{E},\mathbf{H})$ instead of just $\mathbf{E}$, using the $6\times6$ polarizabilites and Green functions of Ref.~\onlinecite{Lunnemann2013,Kwadrin2014}). As in Ref.~\onlinecite{Kwadrin2014}, our starting point is that the response of a periodic lattice of polarizable point particles illuminated by a plane wave (electric field $\mathbf{E}e^{i\mathbf{k}_{||}\cdot \mathbf{r}}$ with parallel wave vector $\mathbf{k}_\parallel$, plus its Fresnel reflection at the mirror) can be summarized entirely by the response $\mathbf{p}$ of the particle at the lattice origin (dynamic polarizability $\bm{\alpha}$) which reads~\cite{GarciadeAbajo2007,Lunnemann2013}
\begin{equation}
\mathbf{p}\\=\bm{\alpha}\left[
 \mathbf{E}+
\sum_{n\neq 0} \mathbf{G}((\mathbf{R}_n,d),(0,d))e^{i\mathbf{k}_{||}\cdot \mathbf{R}_n}\mathbf{p}
\right]
\end{equation}
so that
\begin{equation}
\mathbf{p}=\frac{1}{\bm{\alpha}^{-1} - \sum_{n\neq 0} \mathbf{G}((\mathbf{R}_n,d),(0,d))e^{i\mathbf{k}_{||}\cdot \mathbf{R}_n}}\mathbf{E}.
\label{eq:invalphaminusG}
\end{equation}
Here, the  real space lattice vectors $\mathbf{R}_n$ are taken parallel to the $(x,y)$-plane, as well as the interface which is at $z=0$.  With $\mathbf{G}(\mathbf{r},\mathbf{r'})$ we denote the dyadic Green function in absence of the lattice, but accounting for the interface, for a single dipole,  explicitly defined in Appendix \ref{sec:green_app}.  The polarizability $\bm{\alpha}$  must be corrected for the presence of the substrate. This means it must be chosen to represent a self-consistent scatterer in presence of the environment specified by  $\mathbf{G}(\mathbf{r},\mathbf{r'})$. Mathematically this means that the polarizability must be corrected for backaction through
$$
\bm{\alpha}^{-1}=\bm{\alpha}_\mathrm{free}^{-1} -\mathbf{G}((0,d),(0,d))
$$
where $\bm{\alpha}_\mathrm{free}^{-1}$ is the dynamic polarizability in free space  and $\mathbf{G}((0,d),(0,d))$ represents the interface-induced backaction.
The dyadic Green function for the case of a single planar interface can be separated into a free space and reflected part~\cite{Novotny}
\begin{equation}
\mathbf{G}(r,r')=\mathbf{G}_\mathrm{free}(r,r')+\mathbf{G}_{\mathrm{refl}}(r,r').
\label{eq:dyadic_Green}
\end{equation}
The free space part is simply the field of a dipole in free space (same index as the medium in which the lattice sits), while $\mathbf{G}_\mathrm{refl}$ is a more complicated function constructed in the angular spectrum representation, and requiring a parallel wave vector integral including the Fresnel coefficients of the interface system, described in detail in Appendix \ref{sec:green_app}. This integral has poor convergence properties that require complex integration methods outlined by Paulus et al.~\cite{Paulus2000}. 

Required for evaluating the lattice response through Eq.~(\ref{eq:invalphaminusG})  is the lattice-summed Green function defined through
$$\mathcal{G}^{\neq}:= \sum_{n\neq 0} \mathbf{G}((\mathbf{R}_n,d),(0,d))e^{i\mathbf{k}_{||}\cdot \mathbf{R}_n}$$
where the superscript $\neq$ denotes a sum over all terms $n\neq0$, while lattice sums that do include the $n\neq0$ term are denoted without superscript in this paper. Since the Green function in presence of the single planar interface  separates as a sum, also the lattice-summed Green function can be written as a sum of a free space and a reflected part
\begin{equation}
\mathcal{G}^{\neq}(\mathbf{k}_{\parallel},\mathbf{r})=\mathcal{G}^{\neq}_{\mathrm{free}} (\mathbf{k}_{\parallel},\mathbf{r} )+\mathcal{G}^{\neq}_{\mathrm{refl}} (\mathbf{k}_{\parallel},\mathbf{r}).
\end{equation}
 The lattice summation required to obtain ${\mathcal{G}}_{\mathrm{free}}$  was discussed in detail in previous reports~\cite{Lunnemann2013,Kwadrin2014}. Regarding the reflected part, we find that there is an  intuitive interpretation  in terms of mirror dipoles. If one considers  that  the interface-corrected polarizability is  set by  $
\bm{\alpha}^{-1}=\bm{\alpha}_\mathrm{free}^{-1} -\mathbf{G}((0,d),(0,d))
$, one can rewrite Eq.~(\ref{eq:invalphaminusG}) as 
\begin{equation}
\mathbf{p}=\frac{1}{\bm{\alpha}_\mathrm{free}^{-1} - \tilde{\mathcal{G}}(\mathbf{k}_{||},\mathbf{r})}\mathbf{E}
\label{eq:renorm}
\end{equation}
with
\begin{equation}
\tilde{\mathcal{G}}(\mathbf{k}_{\parallel},\mathbf{r})={\mathcal{G}}^{\neq}_{\mathrm{free}} (\mathbf{k}_{\parallel},\mathbf{r} )+{\mathcal{G}}_{\mathrm{refl}} (\mathbf{k}_{\parallel},\mathbf{r}).
\end{equation}
In terms of mirror image dipoles, the interpretation is that a given particle interacts with all the particles \emph{except} itself in the physical lattice (captured in $\mathcal{G}^{\neq}_{\mathrm{free}}$), and furthermore with all the mirror imaged scatterers \emph{including} its own mirror image (captured in ${\mathcal{G}}_{\mathrm{refl}}$). 

As regards rigorous evaluation of the reflected lattice sum,
\begin{eqnarray}
{\mathcal{G}}_{\mathrm{refl}} (\mathbf{k}_{\parallel},\mathbf{r})& = &
\sum_{m,n} \mathbf{G}_{\mathrm{refl}}(\mathbf{R}_{mn} -\mathbf{r}_{\parallel}) e^{i\mathbf{k}_{\parallel}\cdot \mathbf{R}_{mn}}  
\label{eq:Grefl}
\end{eqnarray}
the problem looks quite formidable since the reflected Green function $\mathbf{G}_{\mathrm{refl}}(\mathbf{r},\mathbf{r'})$ that one needs to sum  has the following mathematical structure, at distance $d$ from the interface
$$
\frac{i}{2\pi} \left[ \int d\mathbf{q}\frac{1}{k_z} \mathbf{M}_\mathrm{refl}e^{i\mathbf{q}\cdot (\mathbf{r}-\mathbf{r'})} e^{ik_z |z+z'|}\right]
$$
Here $\mathbf{M}_\mathrm{refl}$ depends on in-plane wavevector $\mathbf{q}$ and contains the Fresnel coefficients of the interface~\cite{Kwadrin2013PRB}, specified in Appendix \ref{sec:green_app}. The Sommerfeld integral is subtle to perform, as a pole occurs at the light line due to the $1/k_z$ term (where $k_z=\sqrt{\omega^2-|\mathbf{q}|^2}$ represents out-of-plane wave vector), as well as at any wave vector $\mathbf{q}$ that corresponds to a guided mode at the interface.  While summing an infinite set of Sommerfeld integrals appears daunting, in fact the completeness relation 
\begin{equation}
\sum_{m,n}  e^{i\mathbf{k}_{\parallel}\cdot \mathbf{R}_{mn}} =
 \frac{(2\pi)^2}{a^2}\sum_{m,n}
 \delta(\mathbf{k}_{\parallel}-\mathbf{g}_{mn}),
\end{equation}
leads to a remarkable simplification. 
Indeed substituting the completeness relation removes the Sommerfeld integration altogether, instead resulting in a quickly converging sum.  Thus, owing to parallel wave vector conservation, the interface lattice sum is \emph{easier} to calculate than the interface Green function $\mathbf{G}_{\mathrm{refl}}$ itself. The sum reads (with $k^\mathbf{g}_z = \sqrt{k^2-|\mathbf{k}_{||}+\mathbf{g}|^2}$)
\begin{eqnarray}
{\mathcal{G}}_{\mathrm{refl}} (\mathbf{k}_{\parallel},\mathbf{r})& = &
\frac{2i\pi}{a^2} \sum_{m,n} \frac{1}{k^\mathbf{g}_z} \mathbf{M}_{\mathrm{refl}}e^{i(\mathbf{k}_{\parallel}+\mathbf{g})\cdot \mathbf{r}_{\parallel}} e^{ik^\mathbf{g}_z|z+2d|}.
\end{eqnarray}
This sum runs over propagating as well as evanescent diffraction orders, meaning that also any diffractive coupling to guided modes of the underlying mirror are accounted for, if relevant. 

Far-field observables, i.e., reflection, transmission and diffraction efficiencies follow simply once the induced polarizability has been solved for, by using the far field asymptotic expansion of the reflected part of the Green function (Chapter 10 in Ref.~\onlinecite{Novotny}).  As in Ref.~\onlinecite{Kwadrin2014}, the far-field response consists of the specularly reflected beam as well as diffracted orders in the upper and the lower medium.   We note that none of this analysis was specific to a substrate performing as a good mirror. In fact, this formalism can deal with lattices on arbitrary metallic or dielectric stratified media, where the lattices can be either diffractive or subdiffractive. The model also directly applies to, for instance, analyzing the strength of diffractive surface lattice resonances of lattices near the interface of two transparent media, as analyzed by Augui\'e et al.~\cite{Auguie2010}. Arbitrary stratified systems can be included by replacing the Fresnel coefficients implicitly appearing in the reflected Green function by appropriate multilayer Fresnel coefficients.

\section{Results of the full theory}
\subsection{Reflectivity}

\begin{figure}
\centering
\includegraphics[width=0.5\textwidth]{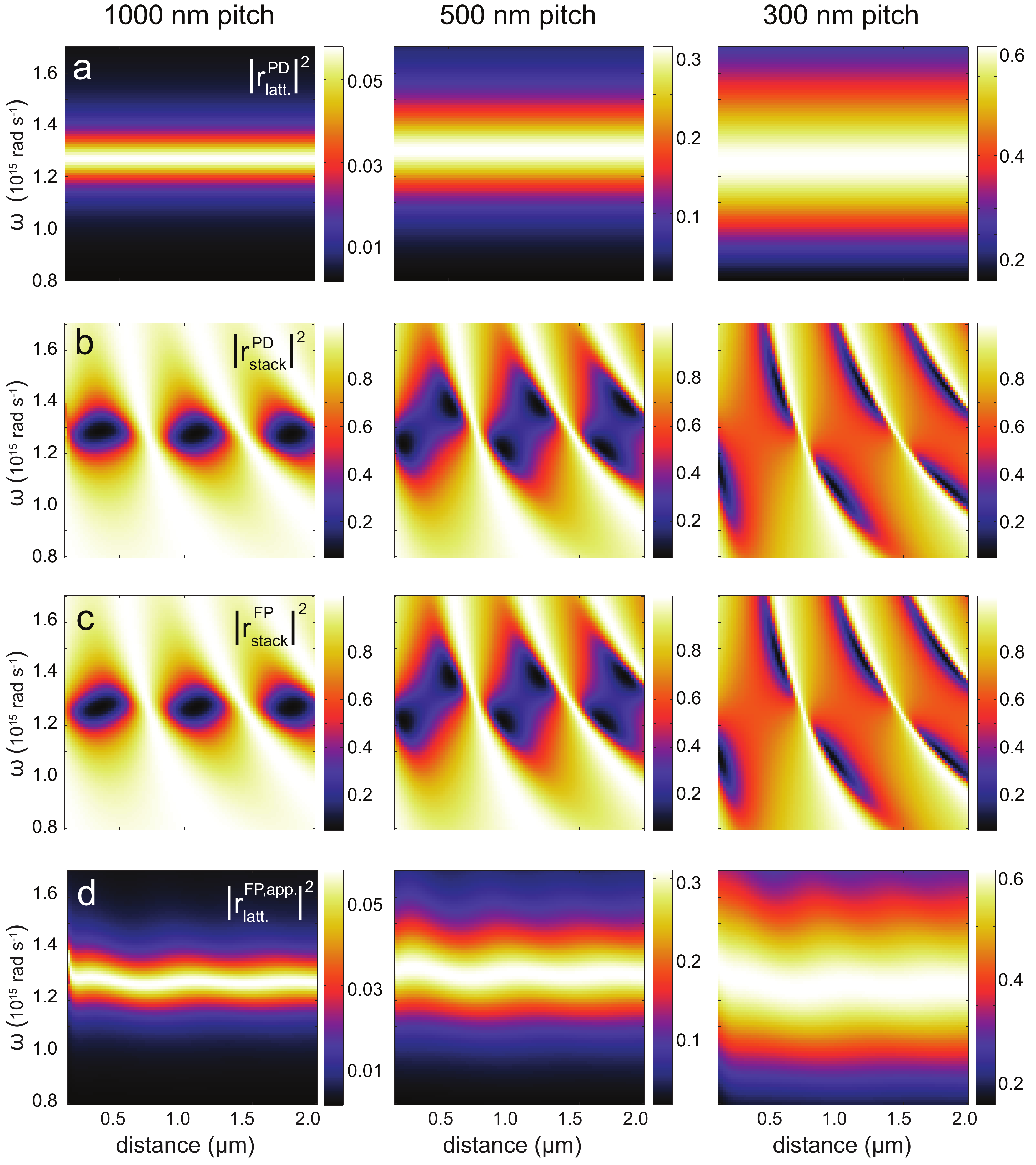}
	\caption{The reflection coefficient for lattices of plasmon bars ($1000, 500, 300\,$nm pitch going from left to right) from full lattice-sum point-dipole calculations.  (a) Lattice reflectivity in absence of any mirror (thereby independent of distance), (b) \emph{full} reflectivity of the system calculated in presence of the mirror, (c) approximate etalon reflectivity calculated by inserting the result of (a) in the Fabry-Perot equation, and (d) apparent lattice reflectivity retrieved by inverting the Fabry-Perot equation taking as input the mirror Fresnel coefficient and the full reflectivity shown in (b).}
\label{fig3:fullreflectivity}
\end{figure}
 
Here we discuss reflection coefficients calculated through the exact Green function formalism set up in the previous section. For comparison, Fig.~\ref{fig3:fullreflectivity} shows the absolute squared of the frequency dependent reflection coefficient  $r^{\mathrm{PD}}_{\mathbf{latt}}$  from point dipole (PD) lattice-sum calculations in \emph{absence} of the mirror, Fig.~\ref{fig3:fullreflectivity} (a), and calculations that contain the mirror, Fig.~\ref{fig3:fullreflectivity} (b), as in the preliminary Fabry-Perot analysis. 

For relatively dilute lattices, e.g., pitch $500\,$nm,  the absolute squared value of the reflection amplitude  $|r^{\mathrm{PD}}_{\mathbf{latt}}|^2$ is small in the whole frequency range, Fig.~\ref{fig3:fullreflectivity} (a). For the full system including mirror, the  off-resonant  $|r^{\mathrm{PD}}_{\mathbf{lattice}}|^2$ is almost $100\,\%$ throughout the distance range as expected since at these frequencies  the response is dominated by the reflection of the silver mirror, Fig.~\ref{fig3:fullreflectivity} (b). On-resonance, we find a low reflectivity of less than $10\,\%$  in minima at distinct mirror-metasurface distances. The combined system has a remarkably \textit{large} reflectivity contrast for the reported  frequency range of $\left[1\%,95\%\right]$, even though  the individual interfaces both feature a \textit{small} reflectivity contrast of $\left[0.5\%,5.5\%\right]$ for the lattice and $\left[98\%,99\%\right]$ for the mirror. The fact that fringes occur at  $\lambda/2$  intervals is consistent with the idea of a Fabry-Perot resonator governed by constructive interferences when the round trip accumulates a full wavelength ($2kd=m2\pi$). At all densities, a fringe of perfect reflectivity occurs when the lattice is right in the node of the standing wave that forms at the mirror.

For  more dense lattices, the etalon reflectivity transits into sharp dispersive features in the reflection coefficient. For the lattice with a pitch of $300\,$nm, the reflectivity is qualitatively very different, having gone from absorptive features on a high background to asymmetric dispersive lineshapes, Fig.~\ref{fig3:fullreflectivity} (b)  right column. The reversal of contrast at $1.2\times10^{15}\,$rad\,s$^{-1}$, i.e., whether the reflectivity is low or high to the left, or the right of the Fabry-Perot resonance is due to the lattice resonance, whose response function goes through a $\pi$ phase flip. This transition could be understood as a transition from coherent mixing of a continuum (mirror reflectivity) with a weak absorption resonance (dilute lattices, walking along paths of constant separation), to mixing of a continuum (mirror reflectivity) with a strong resonance (lattice response). Thus the lineshape change is akin to that in the families of lineshapes of the Fano interference effect that has attracted significant attention in many communities, including recently that of optical scattering~\cite{KivsharRevModPhys}.

 \subsection{Fabry-Perot model and backaction effect on reflectivity}
We find that the Fabry-Perot model is remarkably effective for predicting the reflectivity of the metasurface etalon on basis of just the mirror reflectivity, and the metasurface normal-incidence reflectivity. To bring this out, we compare the full model (i.e., including all particle-particle interactions, including those through the interface),  with the Fabry-Perot model that takes as only input the mirror reflectivity (in absence of the lattice), and the \emph{bare} lattice reflectivity  (i.e., reflectivity calculated with the point dipole method in absence of the substrate).  The resulting $|r^{\mathrm{FP}}_\mathrm{stack}|^2$ is shown as panel (c) for each lattice density in Fig.~\ref{fig3:fullreflectivity}. The same three regimes of reflectivity depending on the strength of the metasurface response are seen, with very similar magnitudes of reflectivity features.  Inspection shows that the calculated reflectivity features are qualitatively very similar.  An important result of our work is that it verifies that the Fabry-Perot model is highly appropriate for metasurface etalons.

\begin{figure}
\includegraphics[width=\columnwidth]{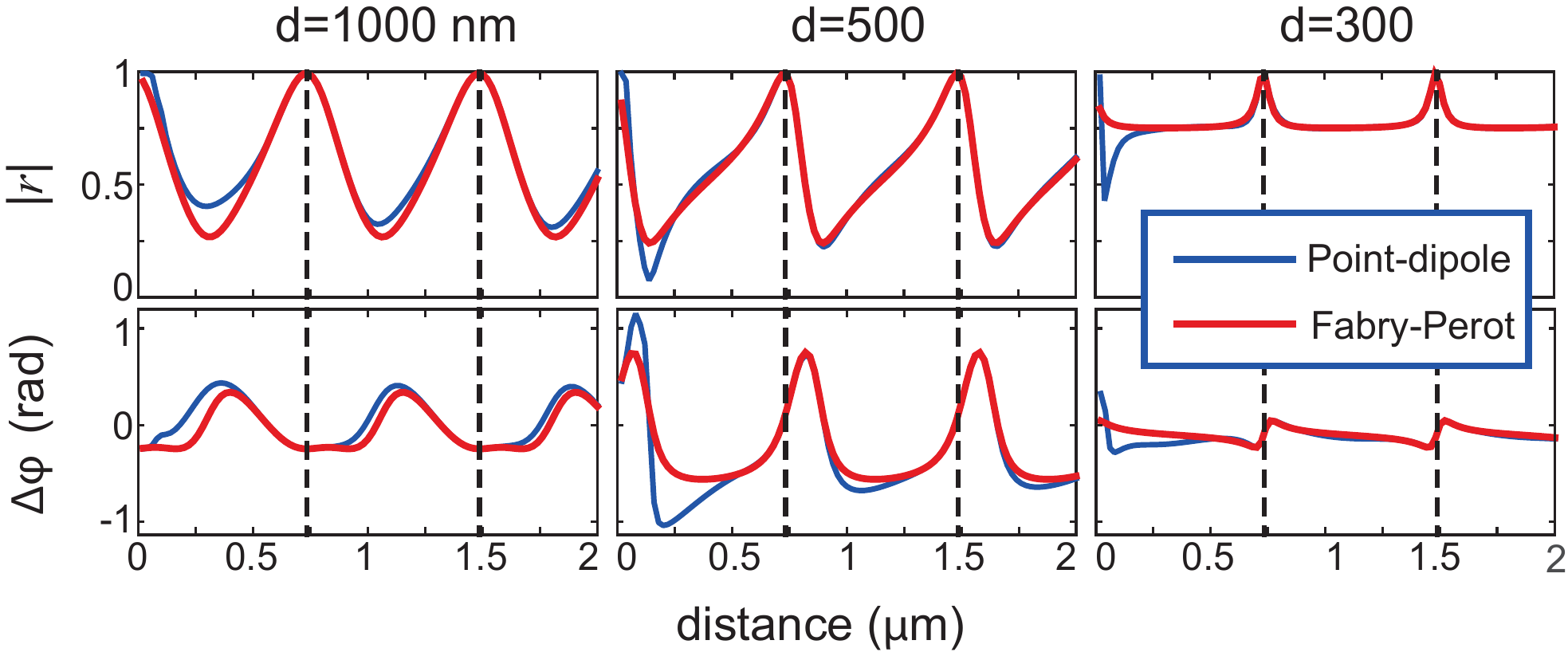}
\caption{Reflection amplitude $|r|$and phase $\Delta\varphi$ versus mirror-metasurface spacing for different lattice densities, at frequency $\omega=\omega_0$.  Blue curves show the full point dipole result, while red curves show the Fabry-Perot result. It should be noted that to focus on the \textit{differences} between the model, the plotted phase is relative to a subtracted trivial linear behavior (due to the fact that in scanning distance, the front reflective surface approaches the observer). 
\label{fig3extra:deviation}}
\end{figure}

More thorough analysis in fact does show deviations. For instance, the precise location of troughs in the reflectivity shifts  as much as $100$ nm in distance  and  up to $5$\%  in frequency, with differences of order $0.1$ in reflectivity contrast.  A clearer representation is offered by constant-frequency line-cuts  that show the reflection amplitude, and the reflection phase in both models, Figure~\ref{fig3extra:deviation} shows such cuts at $\omega=\omega_0$. At pitches of $500$  ($\lambda/3$) and $1000$ nm ($2\lambda/3$) (not diffractive but not deeply subwavelength) the reflection amplitude differs by up to $0.1$,  while the reflection phase differs by up to 0.5 radians (about $30$$^\circ$). This would constitute a significant difference when using the stack as a phase-control element. Comparing line cuts at various frequencies, it is clear that the agreement is worst (i) with increasing pitch, being noticeable for pitches  $\gtrsim \lambda/5$, (ii) closer to the mirror, and (iii) in the low-reflectivity areas, as expected given Eq.~(\ref{eq:Grefl}).  Deviations from the Fabry-Perot model are due to evanescent diffractive contributions, which require significant $e^{2ik_z^\mathbf{g}d}$, favoring small mirror-lattice separation $d$, and large pitch (reducing $ik_z^\mathbf{g}$).  At low reflectivity the lattice is most strongly polarized,  as at the lines reach unit efficiency, the lattice is at a field antinode, at which point  (evanescent) diffractive corrections are irrelevant. The reasoning in terms of evanescent diffractive orders is similar to that by Zhao et al.\cite{ZhaoAdvMat}, where it was found that a similar criterion determines when stacked metamaterial layers are sensitive, or immune to lateral alignment errors.

One viewpoint on the deviations between Fabry-Perot model and the model including all particle and mirror interactions,  is that the metasurface reflectivity is apparently \emph{not} independent of environment.  A fundamental tenet of the standard treatment of dielectric multilayers is that the composite reflectivity can be expressed rigorously in terms of just  Fresnel coefficients for reflection and transmission at each interface and  phase terms associated with propagation across the thickness of each layer. As example, in the two-layer formula Eq.~(\ref{eq:FPmodel}), the coefficients $(r_{nm},t_{nm})$ are  independent of the spacing $d$. The coefficients are determined for each boundary by boundary condition matching. Since a metasurface is not a microscopically smooth  interface to which boundary condition matching applies the assumption that $r_{1,2}$ is independent of the substrate might not hold.   We try then to determine   what is the \textit{apparent} lattice reflectivity that one needs to insert in the Fabry-Perot model to make it match the full calculation.   This is motivated by the use of \emph{retrieval procedures} in metamaterial research, that aim to retrieve effective medium parameters by  inverting experimental observations. Similarly, given the mirror reflectivity, and the mirror-to-lattice spacing ($d$ and $r_{2,3}$), one can envision a retrieval procedure to determine the apparent metasurface reflectivity from the measured complex etalon-reflectivity, which can be  measures using an interferometer. Equation~\ref{eq:FPmodel} can be inverted by  simple algebra to obtain a complex effective $r_{1,2}$ with no ambiguity for each separation $d$.  Figure~\ref{fig3:fullreflectivity}, panel (d) shows the apparent lattice reflectivity for each lattice density. Clearly the retrieval is successful, in that the retrieved apparent reflectivity does not show the very strong fringing contrast of the composite. Nonetheless the apparent reflectivity is not identical to the  reflectivity of the lattice in absence of the mirror. Instead it 
\textit{does} depend on the spacing $d$ between mirror and lattice. The resonance frequency oscillates with distance to the mirror, as does the width. 
  
There are two ways to understand the dependance of the effective reflectivity of the lattice on distance to the interface. In grating terminology, slicing a structure up in layers is only a valid approach if one includes sufficient evanescent diffracted orders to capture near field coupling. It is evident that a Fabry-Perot model that takes only the reflectivity and transmission of propagating orders does not include the evanescent orders, making it fundamentally impossible to predict the response of metasurface stacks using a simple Fabry-Perot model.   As a second way to understand the variation of the apparent lattice reflectivity with the external geometry, we propose that it is akin to backaction correction to the response of a scatterer.  Since the experiments by Drexhage in the 1960's, it is well known that the radiative decay rate of quantum emitters, as well as the radiative damping rate of classical dipoles, depends on distance to a mirror, because interactions of the dipole with its own mirror image cause super- or subradiant damping.  It has   been realized since an experiment by Buchler et al.~\cite{Buchler2005} that for single plasmon particles this means that the scattering cross section varies in strength, width, and resonance center frequency  as the particle is held in front of a mirror (oscillation period around $\lambda/2$).  Here a similar back action effect occurs, however now in form of a lattice of dipoles interacting with an entire lattice of image dipoles.

\subsection{Backaction in polarizability}
\begin{figure}
\centering
\includegraphics[width=0.5\textwidth]{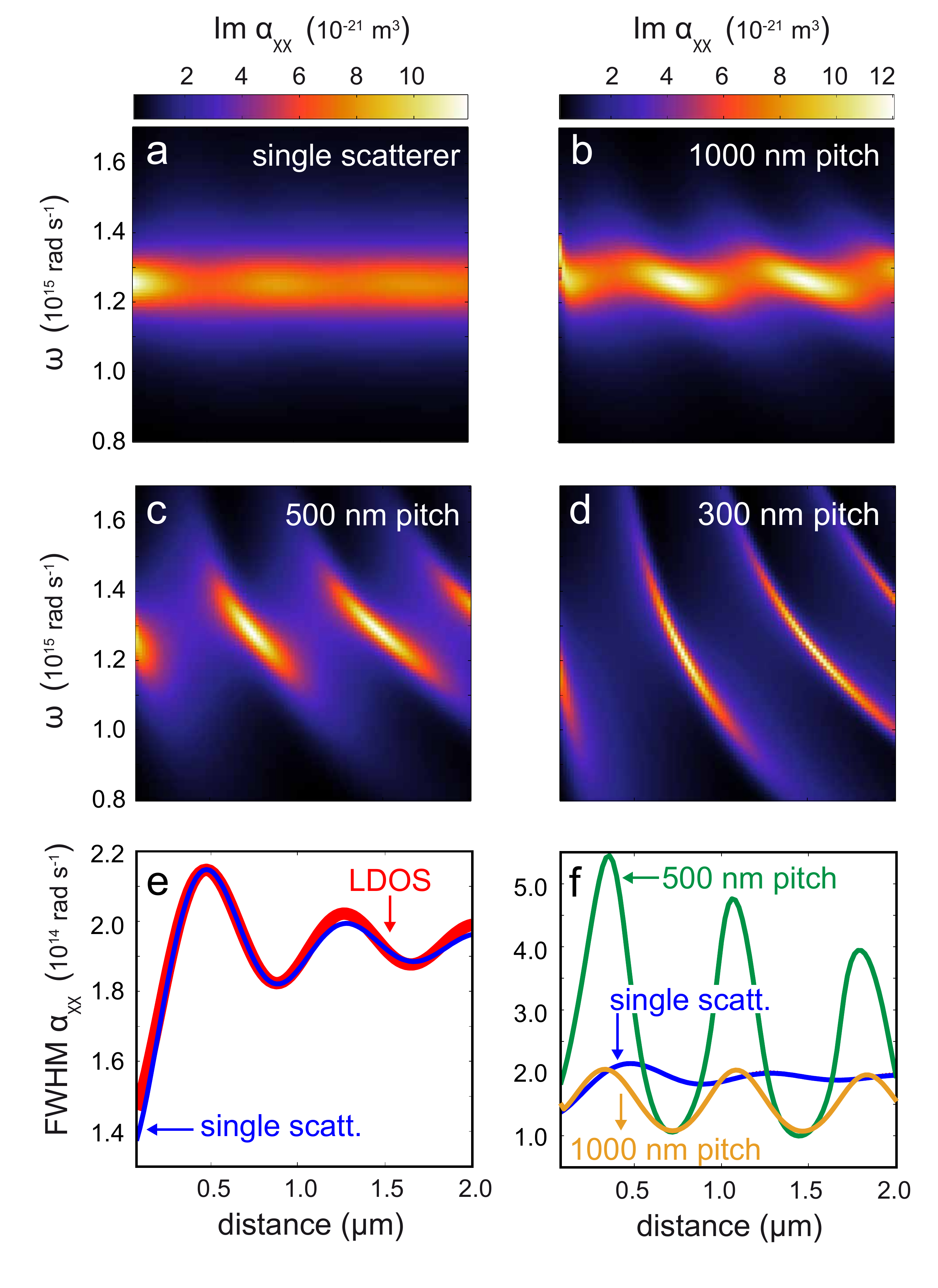}
	\caption{Point dipole lattice-sum calculations to retrieve the polarizability of nanorod scatterers at a distance from a silver interface driven by an $x$-polarized plane wave under normal incidence. The single-particle polarizability (a) varies with distance due to the interaction with its mirror image. Lattice arrangements of $1000\,$nm (b), $500\,$nm (c), and $300\,$nm (d) pitch reveal the increased dispersivenes. The single-particle polarizability linewidth follows the local density of states lineshape for the vac/silver interface (e). For the single scatterer and two most dilute lattice cases it is feasible to compare the single particle linewidth to the overall lattice polarizability linewidth. For $1000\,$nm pitch, the single particle contribution is of comparable strength, while for $500\,$nm pitch, the renormalization effect stemming from the lattice dominates the linewidth (f).}
\label{fig4:PDcalculation}
\end{figure}

Besides analyzing back action effects in the observable, i.e., reflectivity, it is possible to determine how strongly the polarizability of particles is renormalized. Underlying the variation in scattering cross section of a single particle as it is held in front of a reflective surface, is the notion that its polarizability $\alpha$ is renormalized by the interaction with the mirror.   The polarizability quantifies the strength of the induced dipole moment \emph{given} that the incident field applies a set strength of $1$~V/m.   In other words, if one has a single particle in front of a mirror, then upon driving by an external field $\mathbf{E_0}$ (in itself satisfying Maxwell's equations in presence of the mirror, but absence of the particle), the expected dipole response is
$$
\mathbf{p}=\mathbf{\alpha}[\mathbf{E_0} + \mathbf{G}_\mathrm{refl}(\mathbf{r},\mathbf{r})\mathbf{p}] $$
If one defines the polarizability as the response to the externally applied driving  field $ \mathbf{E_0}$, rather then the total field, the consequence is that the polarizability is renormalized
$$\quad \mathbf{p}=\alpha_\mathrm{eff} \mathbf{E}_0$$ 
with
$$\frac{1}{\alpha_\mathrm{eff}} = \frac{1}{{\alpha}_\mathrm{free} }-  \mathbf{G}_\mathrm{refl}(\mathbf{r},\mathbf{r}).$$ 
In a similar way, Eq. (\ref{eq:renorm}) shows that the per-particle polarizability --- when defined as the induced dipole strength in response to the \textit{incident} (as opposed to total) field, is renormalized by the direct interactions with all other particles, and through all the interactions that take place via the mirror. Through the help of Fig.~\ref{fig4:PDcalculation}, we examine each of these renormalization steps.

 Figure \ref{fig4:PDcalculation} shows the frequency and distance-dependent imaginary part of the dynamic polarizability for a single scatterer and lattices with different pitch in front of the silver mirror. We focus on the imaginary part of the polarizability, because in free space $4\pi k\,\mathrm{Im (\alpha)}$ is associated to the extinction cross section. For the bare scatterer $\mathrm{Im (\alpha)}$   presents a Lorentzian peak at $1500\,$nm  with  peak extinction of $\approx 0.3\,\upmu$m$^2$ corresponding to an archetypal plasmonic scatterer~\cite{Husnik}. Once placed in front of the mirror  the  single particle polarizability presents a clearly defined resonance with resonance width, center frequency and strength that vary with distance due to backaction (Fig.\ref{fig4:PDcalculation} (a)). For very small separations less than $100\,$nm, the resonances strongly shift when approaching the mirror. However, since here the dipole approximation is not valid we omit this range from our plots.    The   single-particle polarizability linewidth extracted from Lorentzian fits at each separation follows the  local density of states $\mathrm{Im} \mathbf{G}(\mathbf{r},\mathbf{r})$  modulation associated with the vacuum/silver interface, Fig.~\ref{fig4:PDcalculation} (e).

We now turn to the effective polarizability per object for lattices ranging from dilute to dense ($1000\,$nm to $300\,$nm pitch), Fig.~\ref{fig4:PDcalculation} (b)-(d). The polarizability reduces strongly and broadens significantly with increasing density when considering lattices in absence of the mirror (result not shown).  This broadening with increasing density  is well known for metamaterial lattices~\cite{Linden2004,Sersic2009}, and is due to the fact that many dipoles oscillating in phase ($k_\parallel=0$) experience superradiant damping (radiated power of $N$ dipoles being proportional to $N^2$). Another viewpoint is that the extinction cross section $4\pi k\mathrm{Im (\alpha)}$ per object must remain below the unit cell area. Once placed in front of the mirror, we note that when going from dilute to dense lattices the polarizability is increasingly dominated by a strong feature that is neither solely attributable to the interface nor to the lattice. The weak fringes in polarizability for the single particle sharpen considerably,  and  become very strongly dispersive, forming features along lines of constant  $d/\lambda$. The oscillation period reduces to $\approx 750\,$nm, i.e., exactly  half the single-particle resonance wavelength. For the densest lattices the fringes essentially span over the entire frequency axis. This notion is consistent with the fact that for dense bare lattices, the single particle polarizability looses relevance, as the polarizability is increasingly determined by just the unit cell area. For such dense lattices, the calculated effective polarizability in effect implies that a strong dipole can be induced only whenever the lattice and the mirror satisfy a Fabry-Perot resonance condition. For these conditions, the areal density of maximum achievable polarization is high, providing a clear indication that conditions of perfect absorption can be associated with a strong increase in particle polarizability.

The bottom row of Fig.~\ref{fig4:PDcalculation} shows the FWHM distance dependence for the lattice and interface normalized polarizability. The single-particle polarizability renormalized for the presence of the interface from Fig.~\ref{fig4:PDcalculation} (e) serves as a reference in Fig.~\ref{fig4:PDcalculation} (f).  For dilute lattices, $1000\,$nm   pitch, Fig.~\ref{fig4:PDcalculation} (f) shows that the interface and lattice contribution to the linewidth are comparable. For denser lattices the renormalization effects due to the interaction of the lattice of dipoles with its own mirror image is very large, resulting in a modulation of the polarizability resonance bandwidth by a factor 5 (as opposed to the 25\% effect expected for just a single scatterer at the same mirror). In contrast to the case of a single particle, this modification has a period exactly $\lambda/2$, and a modulation depth that stays much stronger for larger distances, indicative for the fact that backaction is now through an entire extended lattice of mirror dipoles. Importantly, while a single particle mirror image yields backaction through a spherical wave,  for a  mirror-image of scatterers forming a periodic array, backaction is through a reflected wave at $k_\parallel=0$ stemming from combined and fixed-phase related outgoing and receiving reflected plane waves.

An apparent paradox is that the renormalization of the lattice polarizability through backaction with its own mirror image is very large, while the resulting renormalization of the effective lattice reflection constant is small.  Qualitatively,  the renormalization results in the highest polarizability right when the lattice and its mirror image form a subradiant pair as radiative damping is weakest. This condition is exactly coincident with the incident wave $\mathbf{E}_0$ having a null, meaning that the reflectivity is least sensitive to the lattice right when the polarizability is most strongly renormalized. From this reasoning, it is evident that reflection contrast is a poor probe of the large variations in polarizability. While back action effects observed in the far field measure of reflectivity are weak, we expect near field measures such as absorption, nonlinear frequency conversion, or fluorescence spectroscopies to feel the polarizability enhancement much more strongly.

\section{Conclusion}

To conclude, we have derived a calculation method to predict the optical response of metasurfaces and particle gratings held in front of a silver interface. We have applied this model to study the backaction of the interface on the polarizability of the lattice. We conclude that while a Fabry-Perot which takes as input the reflectivities of the substrate and lattice alone provides a reasonably satisfactory description of the joint optical response, in fact there is a surprisingly large effect of the substrate on the particle polarizability. This could significantly impact the result of such composite systems in the near field.

Finally, we note that the calculation method is easily extended to lattices embedded in any arbitrary stratified system. Indeed, for any stratified system the Green function is analytically known and always takes the form of the expression in square brackets in Eq.~\ref{eq:Grefl}, albeit with different tensorial prefactor $\mathbf{M}$. However, the key point is that the difficult parallel wave vector integral always reduces to a discrete sum that is easy to evaluate. Thereby, we expect that this model will be of large utility not just for metasurface physics, but also for predicting the physics of diffractive outcoupling structures patterned on and in LEDs and remote phosphors to improve light generation and directional extraction. Importantly, our model is not restricted to non-diffractive systems, nor to any particular incident $k_\parallel$. Thereby, it contains the full richness of grating diffraction, distributed feedback, and hybridization of lattice modes with the dispersion of any guided modes in the stratified system. 

\begin{acknowledgements}
This work is part of the research program of the Foundation for Fundamental Research on Matter (FOM), which is part of the Netherlands Organisation for Scientific Research (NWO). This work is supported by NanoNextNL, a micro- and nanotechnology consortium of the Government of The Netherlands and 130 partners. 
\end{acknowledgements}

\appendix

\section{Illustrative experiment}\label{Experiment}
\begin{figure}
\centering
\includegraphics[width=0.5\textwidth]{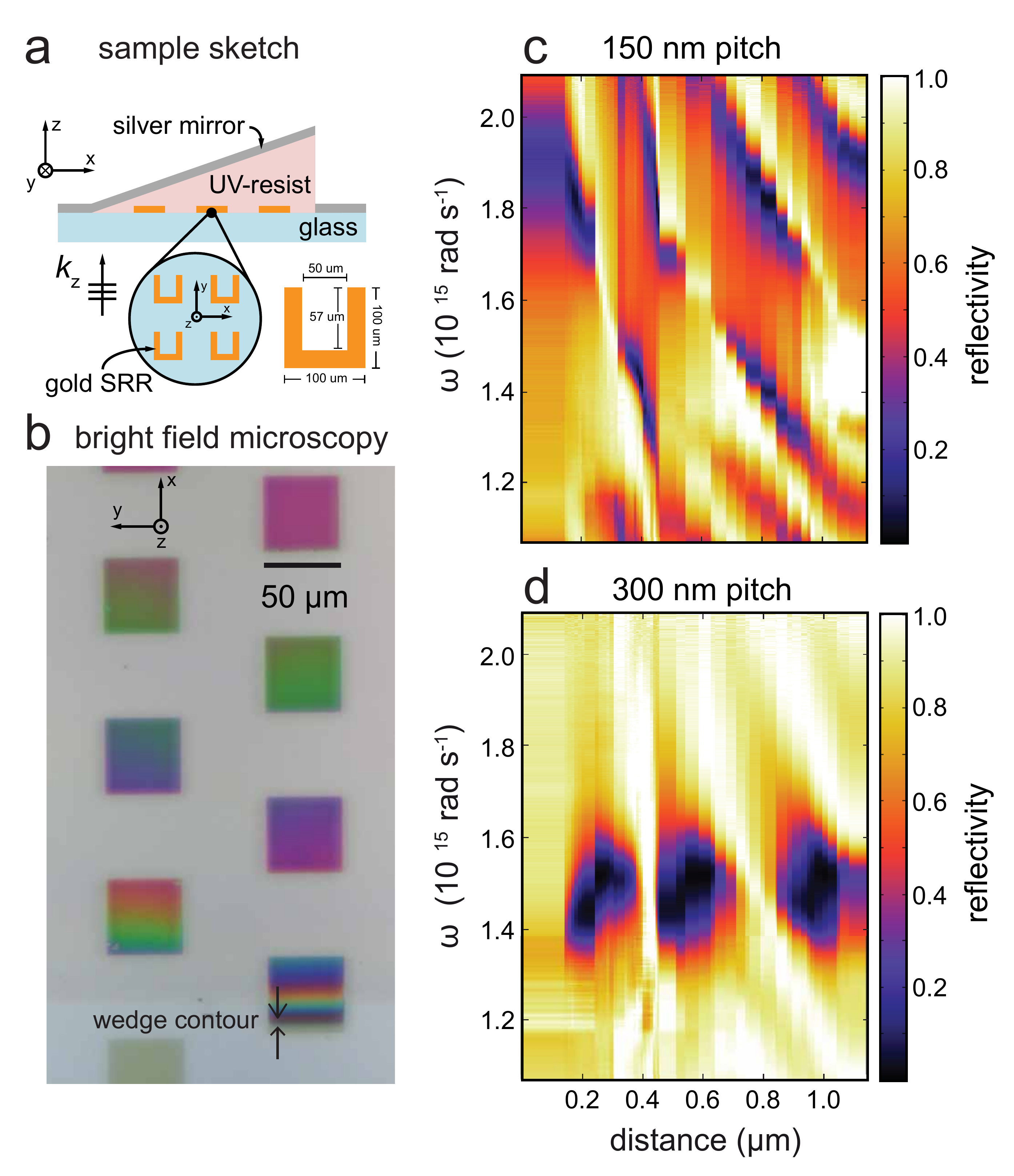}
	\caption{The sample sketch (a) depicts the following sample structure: lattices consisting of gold split ring resonators in a periodic arrangement on glass are separated from a silver mirror by a dielectric wedge (Microposit\texttrademark\  S1813G2 UV-resist) (not to scale). Under bright field illumination (b) these lattices feature a smooth color change in $x$-direction revealing the smooth change in lattice-mirror separation on a $\upmu$m-length scale. (c,d) Experimental reflectivity data for split ring lattices with a pitch of (c) $150\,$nm and (d) $300\,$nm as a function of angular frequency and lattice-mirror separation. }
\label{fig1:sample}
\end{figure}

To illustrate the problem described in this manuscript, we present a simple experiment. We fabricated squared periodic arrays of Au scatterers using electron beam lithography in ZEP-520 resist, thermal evaporation of Au, and lift-off. The arrays were covered by a Microposit\texttrademark\    S1813G2 UV resist wedge by gray-scale lithography, which was used by us earlier as dielectric spacer of smoothly varying height for Drexhage experiments~\cite{Kwadrin2012}. The lattice is located in a quasi-homogeneous environment at a glass ($n=1.52$) - resist ($n=1.6$) \footnote{For the resist, $n$ rises from  $1.59$ to $1.62$ for wavelengths from the telecom infrared range to $\lambda=800$~nm, according to manufacturers and our ellipsometry data} interface. The scatterers are  Au split rings on glass with $100\,$nm arm length, $50 \times 57 \,$nm gap and $30\,$nm height as measured by electron microscopy. While our experiment was originally conceived to probe magnetic back action effects~\cite{Kwadrin2013PRB}, according to our polarizability tensor calculations~\cite{BernalSIE} such small split rings scatter strongly as electric dipoles (polarizable only along the gap, resonant around 1300 nm), with a negligible magnetic response. Thus they are essentially equivalent to plasmon nanorods. The entire structure was coated with $100\,$nm silver that acts as mirror. The sample structure is presented in Fig.~\ref{fig1:sample} (a) together with a bright-field low-magnification microscopy image Fig.~\ref{fig1:sample} (b). In this image, the silver surface appears bright white, while periodic arrays of SRR lattices with $150\,$nm periodicity are visible as darker squares ($50\,\upmu$m e-beam write fields) arranged in a checkerboard pattern.  The millimeter-sized wedge was printed over a macro-array  of split ring arrays, as shown  in Fig.~\ref{fig1:sample} (b). The wedge is faintly visible, it starts at the lower end of the image (indicated by marker) and slopes upwards  along the vertical coordinate (x-direction) in the image, which results in a a smooth color change of the arrays.


We performed normal incidence   reflectivity measurements with a weakly focused beam from a  halogen lamp,  polarized along $x$ to match the split ring fundamental resonance. Given the 5$~\mu$m spot size, we can measure as function of spacer thickness by taking several measurements in a writefield, and furthermore by moving from field to field  along the wedge. White light is spectrally resolved with a fiber coupled NIR-spectrometer (Acton SP-2350i, coupled to an OMA V $1024$ pixel InGaAs array). We plot the measured reflectivity for a very dense ($150\,$nm pitch), and a more dilute ($300\,$nm pitch) lattice, Fig.~\ref{fig1:sample} (c,d).  The bare split ring resonance (no mirror, split ring embedded in the dielectric) is at around $\omega=1.5\times10^{15}\,\mathrm{s}^{-1}$ ($1250\,$nm), which is  consistent with reports by de Hoogh et al.~in Ref. \onlinecite{deHoogh2011}. For the dilute lattice we retrieve a high overall reflectivity, with a set of very deep minima around the single particle resonance. For the dense lattice, the reflectivity shows strong contrast across the entire spectral bandwidth and clearly asymmetric fringes that flip asymmetry at the single particle resonance.  

The standout features of interest for this paper are the following.   First, extremely large reflectivity contrast is obtained with both lattices. Although the reflectivity of the bare lattice is on the order of a few percent  (not shown), with about 30\% single pass absorption, the metasurface etalons show both near zero-percent and near-perfect reflection behavior for particular distances. Second, the two different densities give remarkably different responses. For the more dilute case, the resonance yields pockets of deeply suppressed reflectivity, i.e., strong absorption coincident with the resonance. In contrast in the denser case, the reflectivity is high and featureless right at split ring resonance, but shows deep troughs either just to the red or just to the blue of a geometrical resonance condition (thickness  an integer times  a half-wavelength, meaning that the array is in a node of the standing wave that the incident wave forms at the mirror). The asymmetry flips as one goes through the particle resonance. A somewhat similar phenomenon as we observe for the dense lattice was reported by Ameling et al. for a lattice in a microcavity, who interpreted their observation as an anticrossing of a Fabry-Perot resonance and a resonance of the scattering objects~\cite{Ameling10,AmelingLPRev2013}.

\section{Green's function\label{sec:green_app}}
\subsection{Green's function near a planar interface} 
As described by Eq. (\ref{eq:dyadic_Green}) in Section \ref{sec:full}, the  dyadic Green's function near a planar interface can be separated into a free space (in absence of an interface) and a reflected part $$\mathbf{G}(r,r')=\mathbf{G}_\mathrm{free}(r,r')+\mathbf{G}_{\mathrm{refl}}(r,r').$$
We start from the free space scalar Green function
\begin{equation}
G(\mathbf{r},\mathbf{r}')=\frac{e^{ik|\mathbf{r}-\mathbf{r'}|}}{|\mathbf{r}-\mathbf{r'}|}.\label{eq:scalarG}\end{equation} 
where $k=\omega n/c$ is the wave number in the medium of index $n$ that contains source $\mathbf{r}'$ and observation point $\mathbf{r}$. One finds dyadic Green functions by differentation. The $3\times3$ dyadic for free-space is given by  
\begin{equation}
\mathbf{G}_\mathrm{free}(\mathbf{r},\mathbf{r}') =
\left[\mathbb{I} k^2 +\nabla\nabla \right] G(\mathbf{r},\mathbf{r}')
\label{eq:derive}
\end{equation}   One finds the reflected Green function in three steps. First, one rewrites Eq.~(\ref{eq:scalarG}) through the angular spectrum representation that states
$$
\frac{e^{ik|\mathbf{r}|}}{|\mathbf{r}|} = \frac{i}{2\pi}\int d\mathbf{q}\frac{1}{k_z} e^{i\mathbf{q}\cdot \mathbf{r}_{||}}e^{ik_z|z|}
$$
where $\mathbf{q}$ represents in-plane wave vector, $k_z=\sqrt{k^2-|\mathbf{q}|^2}$ and $\mathbf{r}=(\mathbf{r}_{||},z)$. As second step one applies the differentiation operator (\ref{eq:derive}), while as third step one projects the result in s- and p-polarization, and applies the Fresnel reflection coefficient. The result reads
\begin{multline}
\mathbf{G}_\mathbf{refl}(\mathbf{r},\mathbf{r}') =
\frac{i}{2\pi}\int   
d\mathbf{q}\frac{1}{k_z} 
 e^{i\mathbf{q}\cdot (\mathbf{r}-\mathbf{r'})_{||}}e^{ik_z|z+z'|}\\
\left[\frac{r_s}{(q_x^2+q_y^2)}\mathbf{M}_s -  \frac{r_p}{(q_x^2+q_y^2)}\mathbf{M}_p\right]
\end{multline}
with $r_{s,p}$ of course dependent on $\mathbf{q}$ and
$$
\mathbf{M}_s=k^2
\begin{pmatrix}
 q_y^2   & -q_xq_y    & 0 	\\
-q_xq_y &   q_x^2     &0 	\\
0 & 0 & 0	\\
\end{pmatrix}
$$
and  
$$
\mathbf{M}_p = 
\begin{pmatrix}
q_x^2 k_z^2 &    q_x q_y k_z^2 & { q_x k_z |\mathbf{q}|^2}	\\
q_x q_y k_z^2 &  q_y^2k_z ^2& q_yk_z |\mathbf{q}|^2		  \\
 - q_x k_z |\mathbf{q}|^2 & -  q_yk_z |\mathbf{q}|^2	& -|\mathbf{q}|^4 \\
\end{pmatrix}.
$$


\subsection{Lattice-summed Green function} 

The free space lattice sum is the summation of the dyadic Green function over all real-space lattice sites $\mathbf{R}_{mn}$%
$$
{\mathcal{G}}(\mathbf{k}_{\parallel},\mathbf{r}) = 
\sum_{m,n} \mathbf{G}(\mathbf{R}_{mn} -\mathbf{r}_{\parallel}) e^{i\mathbf{k}_{\parallel}\cdot \mathbf{R}_{mn}},  
$$
which is poorly converging. Following Ref. \onlinecite{Linton2010}  one can split this sum into two convergent terms $\mathcal{G}^{(1)}$ and $\mathcal{G}^{(2)}$, following a technique pioneered by Ewald. Here we  describe the implementation for summing the scalar Green function.  Consider 
\begin{equation}
\Gamma(\bm{k}_{||},\bm{r})=\sum_{m,n}
G(\mathbf{R}_{mn} -\mathbf{r}) e^{i\bm{k}_{||}
\cdot \bm{R}_{mn}}\label{eq:scalLatticeSum}
\end{equation}
with $G(\mathbf{R}_{mn} -\mathbf{r})=G(\mathbf{R}_{mn},\mathbf{r})$ given in Eq.~(\ref{eq:scalarG}). It is possible to write
$$
\Gamma(\bm{k}_{||},\bm{r})=\Gamma^{(1)}(\bm{k}_{||},\bm{r})+\Gamma^{(2)}(\bm{k}_{||},\bm{r})
$$
with
\begin{subequations}
\begin{multline}
\Gamma^{(1)}(\bm{k}_{||},\bm{r})=\frac{\pi}{{\cal{A}}}\sum_{\tilde{m}\tilde{n}}\left\{
\frac{e^{i(\bm{k}_{||}+g_{\tilde{m}\tilde{n}})\cdot
\bm{r}_{||}}}{k^z_{\tilde{m}\tilde{n}}}\right.\\
\cdot\left[
e^{ik^z_{\tilde{m}\tilde{n}}|z|}\mathrm{erfc}\left(\frac{k^z_{\tilde{m}\tilde{n}}}{2\eta}+|z|\eta\right)\right.\\
+
\left.\left. e^{-ik^z_{\tilde{m}\tilde{n}}|z|}\mathrm{erfc}\left(\frac{k^z_{\tilde{m}\tilde{n}}}{2\eta}-|z|\eta\right) \right]\right\}
\label{eq:cylindricalsum}
\end{multline}
and  \ 
\begin{multline}
\Gamma^{(2)}(\bm{k}_{||},\bm{r})=
\sum_{mn}\left\{\frac{e^{i\bm{k}_{||}\cdot\bm{R}_{mn}}}{2\rho_{mn}} 
\cdot\left[ e^{ik\rho_{mn}}\mathrm{erfc}\left(\rho_{mn}\eta
+\frac{ik}{2\eta}\right) \right.\right.\\
+ \left.\left. e^{-ik\rho_{mn}}\mathrm{erfc}\left(\rho_{mn}\eta
-\frac{ik}{2\eta} \right) \right]\right\}.
\label{eq:sphericalsum}
\end{multline} 
\end{subequations}
Here we used $k=\omega/c$, $k^z_{\tilde{m}\tilde{n}}=
\sqrt{k^2-|\bm{k}_{||}+\tensor{g}_{\tilde{m}\tilde{n}}|^2}$,  $\bm{r}=(\bm{r}_{||},z)$, and
$\rho_{mn}=|\bm{R}_{mn}-\bm{r}_{||}|$. Note that the first summation is in reciprocal space, while the second is in real space. The convergence of Eq.~\eqref{eq:sphericalsum} and Eq.~\eqref{eq:cylindricalsum} can be optimized by setting the parameter $\eta$ around $\eta=\sqrt{\pi}/{a}$, where $a$ is the lattice
constant. The limit of the summation over $m$ and $n$ should be at least bigger than the expected number of propagating grating diffraction orders. As discussed in Ref.\cite{Lunnemann2013}, in our metamaterials calculations, with essentially no grating orders, i.e., $ka\leq 2\pi$, we already obtained converged lattice sums for  $|m,n|\leq 5$.   The extension to the required dyadic lattice sum directly follows by applying the differential operator in \eqref{eq:derive}.  Helpful diferentiation rules that exploit the spherical resp. cylindrical coordinate formulation of (\ref{eq:sphericalsum},\ref{eq:cylindricalsum}) are listed in~\onlinecite{Lunnemann2013}.

\bibliographystyle{apsrev4-1}
  \bibliography{KwadrinThesis}

   \end{document}